\documentclass[conference]{IEEEtran}

\usepackage{geometry}
\usepackage{graphicx}
\usepackage{amsfonts} 
\usepackage{amssymb}  
\usepackage{amsmath} 
\usepackage{enumitem}
\usepackage{color}
\usepackage[table]{xcolor}
\usepackage{soul}
\usepackage{url}

\usepackage{multirow}

\usepackage{siunitx}

\newcommand{\rhou}{\rho_{\rm{u}}}
\newcommand{\rhod}{\rho_{\rm{d}}}

\newcommand{\bg}{{\bf{g}}}

\newcommand{\sst}[1]{\mbox{\scriptsize #1}}   

\newcommand{\cC}{\mathbb{C}}

\usepackage{tikz}
\usetikzlibrary{arrows,shapes,snakes,automata,backgrounds,petri}
\usetikzlibrary{positioning}

\newcommand{\neighbors}[1]{\mathcal{N}(#1)}
\newcommand{\neighborsUE}[1]{\mathcal{N}_{\text{UE}}(#1)}
\newcommand{\neighborsAP}[1]{\mathcal{N}_{\text{AP}}(#1)}
\newcommand{\neighborsbullet}[1]{\mathcal{N}_{\bullet}(#1)}
\newcommand{\lin}{\mathcal{L}}
\newcommand{\node}{\pi}

\usepackage[normalem]{ulem}

\begin{document}

\title{A GNN Approach for Cell-Free Massive MIMO}

\author{\IEEEauthorblockN{Lou Sala\"un\IEEEauthorrefmark{1}, Hong Yang\IEEEauthorrefmark{2}, Shashwat Mishra\IEEEauthorrefmark{1} and Chung Shue Chen\IEEEauthorrefmark{1}}

\IEEEauthorblockA{\IEEEauthorrefmark{1}Nokia Bell Labs, 1 Route de Villejust,
Nozay, 91620 France,\\
\IEEEauthorrefmark{2}Nokia Bell Labs, 600 Mountain Avenue,
Murray Hill, NJ 07974 USA,\\
Email: \tt lou.salaun@nokia-bell-labs.com,
h.yang@nokia-bell-labs.com,\\ shashwat.mishra@nokia.com, chung\_shue.chen@nokia-bell-labs.com}

}

\maketitle

\begin{abstract}
Beyond 5G wireless technology Cell-Free Massive MIMO (CFmMIMO) downlink relies on carefully designed precoders and power control to attain uniformly high rate coverage. Many such power control problems can be calculated via second order cone programming (SOCP). In practice, several order of magnitude faster numerical procedure is required because power control has to be rapidly updated to adapt to changing channel conditions. We propose a Graph Neural Network (GNN) based solution to replace SOCP. Specifically, we develop a GNN to obtain downlink max-min power control for a CFmMIMO with maximum ratio transmission (MRT) beamforming. We construct a graph representation of the problem that properly captures the dominant dependence relationship between access points (APs) and user equipments (UEs). We exploit a symmetry property, called permutation equivariance, to attain training simplicity and efficiency. Simulation results show the superiority of our approach in terms of computational complexity, scalability and generalizability for different system sizes and deployment scenarios.
\end{abstract}

\begin{IEEEkeywords}
Cell-Free Massive MIMO, max-min power control, graph neural network, MRT, conjugate beamforming.
\end{IEEEkeywords}

\section{Introduction}\label{s:introduction}
Since the seminal paper \cite{marzetta2010}, Massive MIMO (mMIMO) now forms a backbone of 5G physical layer technology. With magnitudes more service antennas at the base stations, high precision beamforming becomes possible and uniform service quality can be made available in a cellular network for the first time \cite{ym2014}, and the spectral efficiency has been greatly increased over 4G. Similar to previous generations of wireless technologies, to go beyond 5G, substantial increase in spectral efficiency is expected for future mobile networks. 

To materially increase the spectral efficiency beyond cellular mMIMO, a cell-free version of mMIMO was first proposed in \cite{ym2013}, and more detailed theoretical investigations were carried out in \cite{ngo2017, elina2017}. Some key challenges in realizing CFmMIMO were outlined in \cite{zhang2020}. Among them is the practical realization of downlink power control. Machine learning approaches to speed up the calculation of power controls are mostly motivated by real-world deployment requirements. Among existing literature, \cite{carmen2019} used deep learning to perform uplink power allocation for sum-rate and max-min rate optimization; \cite{zhang2021,raja2021}  proposed unsupervised deep learning to obtain uplink power control for several optimization objectives. Downlink power control is in general different from the uplink as each access point (AP) is not only subject to total power constraint, but also must judiciously allocate powers to all served user equipments (UEs) for optimal fairness and interference mitigation. \cite{zhao2020} proposed a deep learning method to approximate a high complexity heuristic algorithm for max-min power control. \cite{yan2020globecom, yan2020, salaun2021deep, luo2022} considered neural network approach for the downlink power optimization. In particular, \cite{salaun2021deep} employed a training data augmentation scheme and developed a CNN (convolutional neural network) to obtain a near optimal downlink max-min power control with reasonable complexity; \cite{luo2022} employed a deep reinforcement learning approach with deep deterministic policy gradient algorithm for the problem. It should be noted that the fully connected layers in~\cite{luo2022} can incur large training complexity and thus limit the scalability and generalizability. 

GNN~\cite{scarselli2009} is a relatively recent neural network architecture that has received a lot of attention in the past few years. It associates graph components with sets of features that can be learned through iterative local computations. It can be very effective for node level, edge level, and graph level prediction tasks. In this paper, we develop a GNN to solve the downlink max-min power control problem with MRT precoding. Our contributions are:
\begin{enumerate}
    \item We formulate the power control problem as a node level prediction task and design a graph structure that properly captures the dominant dependence relationship between APs and UEs. In our CNN approach~\cite{salaun2021deep}, a symmetry property of CFmMIMO was utilized to vastly augment the training examples. This augmentation becomes unnecessary now as we recognize some intrinsic properties of the GNN, known as \textit{permutation equivariance}~\cite{keriven2019universal}. It allows us to construct a GNN that matches the symmetric structure of CFmMIMO, which not only greatly simplifies the training process, but also achieves unprecedented accuracy, scalability, and reusability for different system sizes and deployment scenarios.
    \item Our model optimizes directly each user's SINR by backpropagating the gradient of the SINR loss through the GNN during training. This further enables our GNN to produce near-optimal power control directly. In comparison, we used an additional convex optimization in our CNN approach \cite{salaun2021deep}.
    \item We show through numerical simulations that a single trained GNN achieves practically near-optimal performance on a wide range of scenarios, with up to 128 APs, 32 UEs and various deployment morphologies. We compare the complexity of SOCP, CNN and GNN to confirm the advantage of GNN in large systems.
\end{enumerate}

\section{System Model}\label{s:sys_model}
\subsection{Cell-Free Massive MIMO}
We consider a CFmMIMO system \cite{ngo2017,elina2017} with $M$ APs deployed throughout the coverage area. All APs are connected to a central processing unit (CPU) for precoding and decoding processing and for power control coordination. We assume that each AP has one service antenna, and the $M$-AP system serves $K$ single-antenna UEs simultaneously, where $M$ is much larger than $K$. 

The channel matrix between the $M$ AP antennas and the $K$ user antennas is denoted as:
\begin{equation*}
{\bf G}=\left( \bg_1\hspace{.1in} \cdots \hspace{.1in} \bg_K \right)=\left(
\begin{array}{c}
\bar{\bg}_1^{\sst T}\\
\vdots \\
\bar{\bg}_M^{\sst T}
\end{array}\right)\in {\mathbb C}^{M\times K},
\end{equation*}
where $\bg_k \in \cC^{M}$ is the channel vector between the $k$-th user and the $M$ AP antennas, and $\bar{\bg}_m\in \cC^K$ is the channel vector between the $m$-th AP antenna and the $K$ users, where the superscript $^{\sst T}$ denotes the transpose.
The downlink data channel is modeled as:
\begin{equation}
{\bf x} =  {\bf G}^{\sst T}(\sqrt{\rhod}{\bf s}) + {\bf w},  \label{dldc}
\end{equation}
where ${\bf x}\in {\mathbb C}^K$ is the received signal vector at the $K$ user terminals, $\rhod$ is the downlink signal to noise ratio (SNR) for each AP, and ${\bf s}\in {\mathbb C}^M$ is $M$ precoded inputs to the $M$ antenna ports at the $M$ APs, and ${\bf w}\in {\mathbb C}^K$ is a circularly-symmetric Gaussian noise vector. Each AP has a downlink power constraint, which can be specified as:
\begin{equation}
\| {\mathbb E} ({\bf s}^{*\sst T}\odot {\bf s}) \|_\infty\le 1.   \label{dlpc}
\end{equation}
Here, $\mathbb{E}(\cdot)$ is the expectation, the superscript $^*$ denotes the complex conjugate transpose and $\odot$ denotes the element-wise multiplication.

\subsection{Downlink Max-Min SINR with MRT}\label{s:mrt}

The channel between the $m$-th service antenna and the $k$-th user is denoted by: 
\begin{equation}\label{eq:beta_mk}
g_{m,k}=\sqrt{\beta_{m,k}}h_{m,k},
\end{equation}
where $\beta_{m,k}$ models the large-scale fading that accounts for geometric attenuation and shadow fading, and $h_{m,k}$ models the small-scale fading that accounts for random scattering. In a rich scattering propagation environment, the magnitude of the signal typically varies randomly according to the Rayleigh distribution, thus the small-scaling fading $h_{m, k}$ are modeled as circularly symmetric complex Gaussian, independent and identically distributed random variables.

Under these assumptions, and with MMSE (minimum mean square error) channel estimation based on orthogonal uplink pilot sequence, the ergodic downlink effective SINR for the CFmMIMO system with MRT precoding is given by~\cite{elina2017}:
\begin{align}\label{eq:cf_sinr}
\mbox{SINR}_k = \frac{\rhod \left( \sum_{m=1}^M\sqrt{\alpha_{m,k}\eta_{m,k}}\right)^2}{1+\rhod \sum_{m=1}^M\beta_{m,k}\sum_{k'=1}^K\eta_{m,k'}},
\end{align}
where 
\begin{equation}\label{eq:alpha}
\alpha_{m,k}=\frac{\rhou\tau\beta_{m,k}^2}{1+\rhou\tau\beta_{m,k}}.
\end{equation}
$\alpha_{m,k}$ is the mean-square of the channel estimate, and $\rhod$ and $\rhou$ are the normalized downlink and uplink SNR respectively. $\tau$ is the length of the uplink pilot sequence that is used for channel estimation. $\boldsymbol\eta = \{ \eta_{m,k}\}$ is the downlink power control which is subject to per AP power constraint:
\begin{equation}\label{eq:appc}
\sum_{k'=1}^K\eta_{m,k'}\le 1, \quad \forall m.
\end{equation}

The max-min power control optimization problem can be formulated as
\begin{equation*}\tag{$\mathcal{P}$}\label{P}
\begin{aligned}
& \underset{\boldsymbol\eta}{\mbox{max}}\quad\underset{k}{\mbox{min}}
& & \mbox{SINR}_k, \\
& \text{subject to}
& & \sum_{k'=1}^K\eta_{m,k'}\le 1, \quad \forall m, \\
&&& \eta_{m,k'} \geq 0, \quad \forall m,\,k'.
\end{aligned}
\end{equation*}
An optimal solution to problem~\ref{P} can be obtained with SOCP feasibility bisection search \cite{ngo2017,elina2017,yang2018energy}. However, SOCP's computational complexity becomes impractical as $M$ and $K$ increase. For this reason, we develop a GNN that approximates the solution of SOCP but with practical run-times. The GNN takes as input the large scale fading coefficients, denoted by ${\bf B} = \{\beta_{m,k}\}$, and outputs the power control values $\boldsymbol\eta = \{\eta_{m,k}\}$. We will show that a single GNN model can achieve near-optimal performance over a wide range of system sizes and deployment morphologies.

\section{Graph Neural Network-Based Power Control}\label{sec:gnn}

\subsection{Heterogeneous Graph Representation}\label{sec:graph_repr}

We generate datasets composed of optimal $({\bf B}, \boldsymbol\eta)$ pairs, which are used to train and evaluate the neural network. Each dataset corresponds to a different simulation scenario characterized by a triplet $(M, K, mor)$, where $mor \in \{urban, suburban, rural\}$ denotes the deployment and radio propagation morphology. For each example of the dataset, we generate ${\bf B}$ with $M \times K$ coefficients, as described in~\cite{yang2018energy} and following the ITU-R specifications~\cite{ITU-R2009}. The corresponding power control $\boldsymbol\eta$ is obtained by SOCP.

One can notice that, if the AP indices $\{1,\ldots,M\}$ are permuted and/or the UE indices $\{1,\ldots,K\}$ are permuted at the input ${\bf B}$, then the optimal output would be $\boldsymbol\eta$ permuted in the same order. In other words, problem~\ref{P} does not depend on the choice of indices for the UEs and APs. This property is called \textit{permutation equivariance}, and GNNs are known to inherently satisfy permutation invariance and equivariance~\cite{keriven2019universal}.

To apply the GNN on our data, we need to represent them as graphs. We convert each data sample to a directed graph $\mathcal{G} = (V, E)$, where $V$ is the set of nodes and $E$ the set of directed edges. We create one node for each pair $(m, k)\in\{1,\ldots,M\}\times\{1,\ldots,K\}$, for a total of $MK$ nodes. The one-to-one mapping between the $(m,k)$ pair and the node index $i\in V=\{1,\cdots, MK\}$ is denoted by $\node(m,k)=i$. Conversely, we have $\node^{-1}(i) = (m,k)$. Each node $i\in V$ is associated with a tensor $h_i$ called \textit{node feature}. The initial node features are the input $\bf B$ of the problem: for all $i\in V$, $h_i(0) = \beta_{m,k}$, where $\node(m, k) = i$. Then, the GNN updates the node features through $T$ iterations, i.e., $h_i(t)$, for $t = 1,\ldots,T$. The goal is to approximate the optimal output of the problem $\boldsymbol\eta$ with the final features: for all $i\in V$, $h_i(T) \approx \eta_{m,k}$, where $\node(m, k) = i$. The intermediate features, for $t = 1,\ldots,T-1$, are called \textit{hidden features} or \textit{hidden layers}.

\begin{figure}
\centering
\resizebox{0.45\textwidth}{!}{%
\begin{tikzpicture}

  \tikzstyle{node}=[circle,line width=0.5mm,draw=black!80,minimum size=2cm]
	\tikzstyle{ue_node}=[circle,line width=0.5mm,draw=blue!80,fill=blue!20,minimum size=2cm]
	\tikzstyle{ap_node}=[circle,line width=0.5mm,draw=orange!80,fill=orange!20,minimum size=2cm]
	\tikzset{edge/.style = {->}}

  \begin{scope}[xshift=-5cm]
    \node[ue_node,label={[label distance=1cm]above:\Large\textcolor{blue}{$\neighborsUE{i}$}}] (n1)  {$\pi(1,k)$};
		\node[ue_node] (n2) [below=1.2cm of n1]       {$\pi(m\!-\!1,k)$};
		\node[ue_node] (n3) [below=0.2cm of n2]				{$\pi(m\!+\!1,k)$};
		\node[ue_node] (n4) [below=1.2cm of n3]				{$\pi(M,k)$};
		\path (n1) -- node[auto=false, rotate=90]{$\bullet\bullet\bullet$} (n2);
		\path (n3) -- node[auto=false, rotate=90]{$\bullet\bullet\bullet$} (n4);
		\draw[line width=1mm,draw=blue!80,dashed] (0cm,-4.4cm) circle[x radius=2cm, y radius=6cm, very thick];
  \end{scope}

  \begin{scope}[xshift=5cm]
    \node[ap_node,label={[label distance=1cm]above:\Large\textcolor{orange}{$\neighborsAP{i}$}}] (n1')  {$\pi(m,1)$};
		\node[ap_node] (n2') [below=1.2cm of n1']      {$\pi(m,k\!-\!1)$};
		\node[ap_node] (n3') [below=0.2cm of n2']			 {$\pi(m,k\!+\!1)$};
		\node[ap_node] (n4') [below=1.2cm of n3']			 {$\pi(m,K)$};
		\path (n1') -- node[auto=false, rotate=90]{$\bullet\bullet\bullet$} (n2');
		\path (n3') -- node[auto=false, rotate=90]{$\bullet\bullet\bullet$} (n4');
		\draw[line width=1mm,draw=orange!80,dashed] (0cm,-4.4cm) circle[x radius=2cm, y radius=6cm, very thick];
  \end{scope}

  \begin{scope}[yshift=-4.4cm]
	  \node[node] (n)   {$\pi(m,k)=i$};
  \end{scope}
	
	\draw[edge,draw=blue!80,line width=0.75mm] (n) to node[sloped,xshift=5mm,yshift=5mm]{\large\textcolor{blue}{UE type edges}}(n1);
	\draw[edge,draw=blue!80,line width=0.75mm] (n) to node[]{}(n2);
	\draw[edge,draw=blue!80,line width=0.75mm] (n) to node[]{}(n3);
	\draw[edge,draw=blue!80,line width=0.75mm] (n) to node[]{}(n4);
	
	\draw[edge,draw=orange!80,line width=0.75mm] (n) to node[sloped,xshift=-4mm,yshift=5mm]{\large\textcolor{orange}{AP type edges}}(n1');
	\draw[edge,draw=orange!80,line width=0.75mm] (n) to node[]{}(n2');
	\draw[edge,draw=orange!80,line width=0.75mm] (n) to node[]{}(n3');
	\draw[edge,draw=orange!80,line width=0.75mm] (n) to node[]{}(n4');

\end{tikzpicture}
}%
\caption{A typical node $\pi(m,k)=i\in V$ and its two types of neighbors}\label{fig:node_structure}

\end{figure}
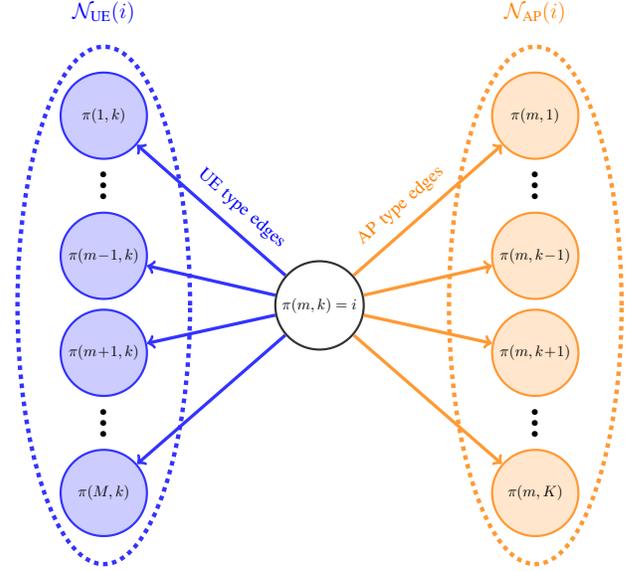

Since the GNN performs local computations on each node and their neighbors connected by an edge, we choose to create one edge between two nodes if they share the same AP or the same UE. There is no self-loop, i.e., $\forall\, i\in V,\, (i, i)\notin E$. Let $e\in E$, we define $\text{type}(e) \in \{\text{AP}, \text{UE}\}$ to be the edge type. Each edge is either of type AP if they share the same AP, or of type UE if they share the same UE. Hence, the edge construction is summarized by the following two statements, for all $m, m'\in\{1,\ldots,M\}$, $m\neq m'$, and $k, k'\in\{1,\ldots,K\}$, $k \neq k'$:
\begin{enumerate}
    \item $e=(\node(m,k),\,\node(m,k'))\in E$ and $\text{type}(e) = \text{AP}$,
    \item $e=(\node(m,k),\,\node(m',k))\in E$ and $\text{type}(e) = \text{UE}$.
\end{enumerate}
We consider heterogeneous edge types so the GNN can distinguish between these relationships and apply different operations on each type of edges.
Let $i\in V$, we define its set of neighbors as:
\begin{equation*}
    \neighbors{i} = \{j\in V \colon (i,j) \in E\}.
\end{equation*}
Similarly, we define the set of neighbors of type UE and type AP as:
\begin{align*}
    \neighborsUE{i} = \{j\in V \colon (i,j) \in E \text{ and } \text{type}(i,j)=\text{UE}\},\\
    \neighborsAP{i} = \{j\in V \colon (i,j) \in E \text{ and } \text{type}(i,j)=\text{AP}\}.
\end{align*}
Fig.~\ref{fig:node_structure} depicts the neighbors of a typical node and its two types of edges. 
For $i\in V$, we have: 
\begin{align*}
    \neighborsUE{i} \cup \neighborsAP{i} &= \neighbors{i}, \\
    \neighborsUE{i} \cap \neighborsAP{i} &= \emptyset.
\end{align*}
Each $\neighborsUE{i}$ has $M-1$ elements, each $\neighborsAP{i}$ has $K-1$ elements, and each $\neighbors{i}$ has $M+K-2$ elements.

This node-edge structure associates nodes with 1st order dependence as neighbors and distinguishes the nodes that share APs and nodes that share UEs.

\subsection{Data Preprocessing}

The graph data preprocessing is done similarly to~\cite{salaun2021deep}. We first apply a $\log_2$ transformation to all the input and output features. This way, the features are within the same order of magnitude, and the GNN is able to extract useful information. As an example, if $\beta_{m,k}$ takes values in $\left[10^{-15},10^{-5}\right]$, then $-50 < \log_2{\beta_{m,k}} < -16$. Then, we normalize the resulting values so that the input features and output features each have zero mean and unit standard deviation over all examples of the dataset. This is a common practice to help speed up the training, as the model does not have to learn the statistics of the data.

\subsection{The Structure of the Neural Network}

In this sub-section, we use notation $\lin$ to denote the linear operation that takes as input a tensor $x$ of size $n$ and outputs a tensor $\lin(x)$ of size $m$ as follows:
\begin{equation*}
\lin(x) = W x + b,    
\end{equation*}
where $W \in {\mathbb R}^{m\times n}$ and $b \in {\mathbb R}^{m}$ are trainable parameters called \textit{weight} and \textit{bias}. Note that, the linear layers used in equations~\eqref{eqn:layer_transformer} and~\eqref{eqn:layer_attention} do not share any trainable parameter. To differentiate them without ambiguity, we write $\lin$ with different subscripts and superscripts.

The feature tensor of each node $i\in V$ is updated based on the features of its direct neighbors at the previous iteration. That is, $h_i(t+1)$ is obtained from $h_i(t)$ and $h_j(t)$, for $j\in\neighbors{i}$. The update rule is as follows:
\begin{equation}\label{eqn:layer}
    h_i(t+1) = \mbox{Norm}\!\left(\mbox{ReLU}\!\left(
    f_{\text{AP}, t}(i)+f_{\text{UE}, t}(i)
    \right)\right),
\end{equation}
where \mbox{Norm} denotes the layer normalization~\cite{ba2016layer}, and \mbox{ReLU} is the rectified linear unit activation function. The functions $ f_{\bullet, t}$, for $\bullet \in \{\text{AP}, \text{UE}\}$, are inspired from the graph transformer of the UniMP model~\cite[Section 3.1]{shi2020masked}, and are defined as:
\begin{align}
    f_{\bullet, t}(i) = \bigoplus_{c=1}^C \bigg(
    &\lin_{\bullet, c, t}^1(h_i(t)) \quad +
    \nonumber\\
    &\sum_{j\in\neighborsbullet{i}} \alpha_{\bullet, c, t}(i, j)\times \lin_{\bullet, c, t}^2(h_j(t))\bigg),\label{eqn:layer_transformer}
\end{align}
where $f_{\text{AP}, t}(i)$ depends on the set of nodes $\neighborsAP{i}\cup \{i\}$, and $f_{\text{UE}, t}(i)$ depends on $\neighborsUE{i}\cup \{i\}$.
We implement each of these two functions with $C=2$ attention heads. The two attention heads corresponding to $c=1$ and $c=2$ are concatenated into a single tensor. This concatenation operation is denoted by $\oplus$. Let $d$ be the tensor size of each attention head, then $f_{\bullet, t}(i)$ is a tensor of size $2d$. Finally, $\alpha_{\bullet, c, t}(i, j)$ is the \mbox{$c$-th} attention coefficient between source node $i$ and destination node $j$, such that:
\begin{equation}\label{eqn:layer_attention}
    \alpha_{\bullet, c, t}(i, j) = \frac{\langle
    \lin_{\bullet, c, t}^3(h_i(t)),
    \lin_{\bullet, c, t}^4(h_j(t))
    \rangle}
    {\sum_{u\in\neighborsbullet{i}}\langle
    \lin_{\bullet, c, t}^3(h_i(t)),
    \lin_{\bullet, c, t}^4(h_u(t))
    \rangle},
\end{equation}
where $\langle x, y \rangle = \exp\left(\frac{x^T y}{\sqrt{d}}\right)$, is the exponential scale dot-product~\cite{vaswani2017attention} and $d$ is the tensor size of each head.

The idea behind the above multi-head attention network is that each node can focus on a subset of its neighbors that are of interest instead of equally considering all the neighbors. 
Indeed, the attention coefficient controls which feature $h_j(t)$ will contribute to $h_i(t+1)$ through Eqn.~\eqref{eqn:layer_transformer}: the feature is discarded if $\alpha_{\bullet, c, t}(i, j) \approx 0$, and kept otherwise.
The level of dependence between two APs (or UEs) varies due to their relative geographic location. We use attention to efficiently learn the complex relationship between nodes. Furthermore, with multiple heads, different attention levels can be assigned to the various features of the same neighbor. Here, we implement $C=2$ attention heads. A further increase in the number of heads does not significantly improve the performance for this problem.
During our study, we observe that the attention mechanism greatly improves the GNN performance and generalizability to large number of APs and UEs.

We can see that whenever the input is permuted, the $\node$ mapping is permuted, but the neighborhood of each $\node(m,k)$ node remains the same. Furthermore, the operations defined above do not depend on any ordering of the neighbors, since the neighboring features $h_j(t)$, for $j\in\neighbors{i}$, are summed in~\eqref{eqn:layer_transformer}. As a consequence, the GNN guarantees that the output is also permuted equivariantly. Thus, the GNN satisfies the permutation equivariance property of our problem~\ref{P}.

By experiments, we optimized the structure of the GNN to achieve near-optimal performance with reasonable complexity. Our GNN model contains 9 hidden layers, i.e., $T = 10$, with the following node feature tensor sizes:
\begin{equation*}
    \text{Layer sizes: } (\text{in}\!=\!1, 8, 8, 16, 16, 32, 16, 16, 8, 8, \text{out}\!=\!1).
\end{equation*}
As expected, the input and output both have a single value per node representing respectively the large scale fading coefficient, and the power control coefficient for each channel. Each hidden layer, for $t = 1,\ldots,9$, is obtained from the previous layer by applying the multi-head attention neural network defined in Eqn.~\eqref{eqn:layer}. The final output tensor is obtained by applying a simple linear activation of the form $h_i(T) = \lin_{\text{out}}(h_i(T-1))$, for all node $i\in V$. To guarantee that the power constraints in~\ref{P} are satisfied, we apply the following two operations on the output tensor:
(i) If any power coefficient is negative, we set it to zero.
(ii) If the power budget constraint is violated for an AP $m$, then we renormalize its transmit powers. That is, if $\sum_{k'=1}^K\eta_{m,k'} > 1$, then for all UE $k$, we assign $\eta_{m,k} \gets \eta_{m,k}/\sum_{k'=1}^K\eta_{m,k'}$.

\subsection{Training and SINR Loss Function}
The training dataset is composed of 4 scenarios: $(M, K, mor) = (32, 6, urban),\allowbreak (32, 9, urban),\allowbreak (64, 9, urban),\allowbreak (64, 18, urban)$. Each scenario has 20,000 samples, for a total of 80,000 training samples. In comparison, the training dataset in~\cite{salaun2021deep} uses $36,000$ raw data for the $(32, 6, urban)$ and $(32, 9, urban)$ scenarios only. These data are then augmented to obtain $\num{2.16e+6}$ training samples. The augmentation consists of duplicating the samples $60$ times and applying random row-permutations and column-permutations to them. This augmentation is needed for the CNN to learn the problem's permutation equivariance and achieve good performance. As explained in the previous sub-section, the GNN inherently satisfies this property. Therefore, such a data augmentation is not needed, and the GNN model converges faster during training. Furthermore, the CNN cannot be extended to larger scenarios without a massive amount of augmented data, as well as an increasing number of trainable parameters to represent the equivariance property. This is a major shortcoming of CNN, which we overcome with the proposed GNN.

The loss function used for training is the mean square error of the per-user SINR, which can be calculated as:
\begin{equation*}
    \frac{1}{K}\sum_{k=1}^K\left(\mbox{SINR}_k-\mbox{SINR}'_k\right)^2,
\end{equation*}
where $\mbox{SINR}_k$ is the optimal SINR of user $k$ obtained by SOCP, and $\mbox{SINR}'_k$ is the SINR computed from the GNN predicted power coefficients $\boldsymbol\eta'$. 
The proposed loss function is differentiable and varies continuously with each user's SINR. By backpropagating the loss through the GNN, it adjusts its hidden layers based on which user's SINR should be increased or decreased at each training step.

We use the Adam optimizer~\cite{kingma2014adam} with a learning rate of~$\num{7e-4}$ to train the GNN model. The batch size is 64, and the training is stopped after 100 epochs.

\section{Numerical Results}\label{s:sim_res}
In this section, we show the performance of our GNN in terms of spectral efficiency, computational complexity, and generalizability for different system sizes and deployment morphologies. We compare it to the results obtained in~\cite{salaun2021deep} for the CNN and the optimal SOCP benchmark. 
To evaluate the complexity of each algorithm, we count their number of floating point operations (FLOPs) during execution. Each multiplication or addition
counts as one FLOP.

The number of APs in our simulations ranges from $M=5$ to $128$. The number of UEs ranges from $K=5$ to $32$. These APs and UEs
are randomly distributed in a circular area within a radius of
500 meters for the urban scenario, 1 km for suburban and 4
km for rural. We consider the "NLoS" propagation model specified in~\cite{ITU-R2009}, and the path loss parameters used are the same as in~\cite{salaun2021deep}.

The figures in this section show the spectral efficiency cumulative distribution function (CDF) achieved by the different algorithms for various simulation scenarios $(M, K, mor)$. We generate 1,000 large-scale fading realizations for each simulation scenario.
The \textit{performance loss at median} refers to the relative difference in spectral efficiency between the deep learning scheme (GNN or CNN) and the optimal solution obtained by SOCP, taken at the median of the CDF. The \textit{95\%-likely performance} refers to the spectral efficiency at the 5-th percentile, i.e., it indicates the coverage quality for 95\% of the users.

All the simulations and algorithms are implemented in Python 3. The GNN is based on the PyTorch Geometric library~\cite{FeyLenssen2019}. The CNN is implemented in TensorFlow, and the SOCP problem~\ref{P} is solved using Mosek~\cite{andersen2003implementing}.

\subsection{Comparison of GNN and CNN}
We compare the performance of GNN and CNN on the scenarios simulated in paper~\cite{salaun2021deep}.
The results are summarized in Table~\ref{tab:gnn_vs_cnn}.
In the urban deployments with 32 APs, GNN achieves less than $0.6\%$ performance loss at median, while CNN has about $2.6\%$ loss. In the suburban and rural scenarios, the performance losses are respectively of $1.55\%$ and $1.35\%$ for the GNN versus $2.81\%$ and $2.84\%$ for the CNN. Besides, we see that the GNN generalizes well when the number of APs changes to 24, achieving similar performance loss of $0.72\%$, while the loss of the CNN degrades significantly to $14.6\%$.

In terms of FLOPs, GNN requires approximately 4 to 9 times more operations than CNN in these small systems. Nevertheless, it is implementable in practice as we observe an execution time of less than 100ms on a CPU\footnote{with the following specifications: Intel Core i5 CPU with 8 GB of RAM, Windows 10, 64 bits.}. We will see in the next subsection that the run-time of GNN remains practical when the number of APs and UEs increases.

\subsection{GNN Spectral Efficiency and Complexity}


\begin{table}[t]
    \centering
    \caption{Comparison of GNN and CNN}\label{tab:gnn_vs_cnn}
    \begin{tabular}{ |c|c|c|c|c| }
         \hline
         \multirow{2}{*}{Scenario} & \multicolumn{2}{c|}{FLOPs \vphantom{$\frac{1}{1}$}} 
         & \multicolumn{2}{c|}{Loss at median} \\ \cline{2-5}
         & GNN \vphantom{$\frac{1}{1}$} & CNN & GNN & CNN \\
      \hline
      Urban & \multirow{2}{*}{\num{1.5e7}} & \multirow{2}{*}{\num{3.7e6}} & \multirow{2}{*}{$0.72\%$} &
      \multirow{2}{*}{$14.60\%$}
      \\ 24 APs, 5 UEs &  & &  & \\
      \hline
      Urban & \multirow{2}{*}{\num{1.9e7}} & \multirow{2}{*}{\num{3.7e6}} & \multirow{2}{*}{$0.48\%$} &
      \multirow{2}{*}{$2.62\%$}
      \\ 32 APs, 5 UEs &  & &  & \\
      \hline
      Urban & \multirow{2}{*}{\num{3.2e7}} & \multirow{2}{*}{\num{3.7e6}} & \multirow{2}{*}{$0.58\%$} &
      \multirow{2}{*}{$2.68\%$}
      \\ 32 APs, 9 UEs &  & &  & \\
      \hline      
      Suburban & \multirow{2}{*}{\num{3.2e7}} & \multirow{2}{*}{\num{3.7e6}} & \multirow{2}{*}{$1.55\%$} &
      \multirow{2}{*}{$2.81\%$}
      \\ 32 APs, 9 UEs &  & &  & \\
      \hline
      Rural & \multirow{2}{*}{\num{3.2e7}} & \multirow{2}{*}{\num{3.7e6}} & \multirow{2}{*}{$1.35\%$} &
      \multirow{2}{*}{$2.84\%$}
      \\ 32 APs, 9 UEs &  & &  & \\
      \hline
    \end{tabular}
    \vspace{-0.5cm}
\end{table}

\begin{figure}
\centering
\includegraphics[width=\linewidth]{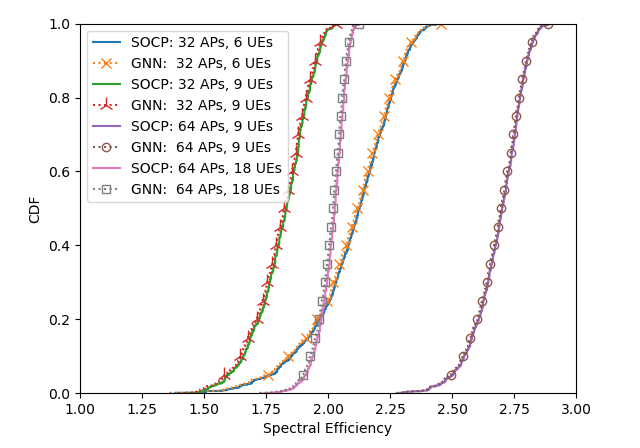}
\caption{Spectral efficiency on urban scenarios}\label{fig:se_urban}
\vspace{-0.4cm}
\end{figure}

Fig.~\ref{fig:se_urban} shows the spectral efficiency of GNN and SOCP on urban scenarios similar to the ones used for training. GNN reaches unprecedented accuracy, with less than $0.5\%$ performance loss at median in all 4 scenarios. Moreover, the 95\%-likely spectral efficiency is at most $0.017$ bits/s/Hz away from optimal. This demonstrates the relevance of our graph representation for problem~\ref{P}, and GNN's effectiveness in learning the optimal power control. Besides the figures, the simulation results of this subsection are summarized in Table~\ref{tab:flops_count}.

\begin{figure}
\centering
\includegraphics[width=\linewidth]{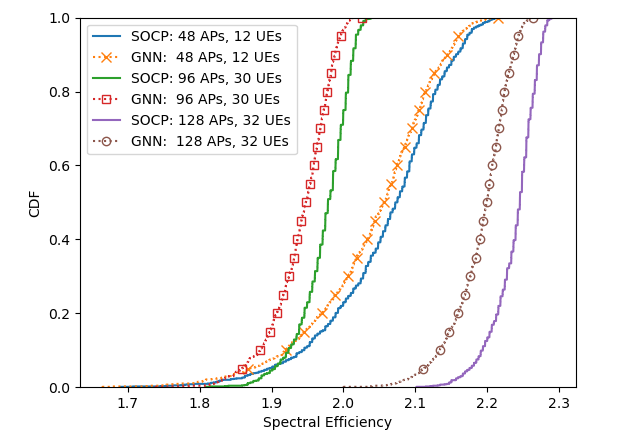}
\caption{Spectral efficiency on urban scenarios with graph sizes (number of UEs and APs) that have not been seen by the GNN during training}\label{fig:se_urban_untrained}
\vspace{-0.5cm}
\end{figure}

To see how our model generalizes to different graph sizes, we run simulations for $(M, K, mor) = (48, 12, urban)$, $(96, 30, urban)$ and $(128, 32, urban)$, see Fig.~\ref{fig:se_urban_untrained}. Their performance loss at median are respectively $0.77\%$, $1.56\%$ and $2.05\%$. We observe that the relative performance decreases slightly as the graph size increases. Overall, the performance remains very competitive even for graphs that are twice larger than the graphs used for training, e.g., 128 APs and 32 UEs. 

We also validate our GNN model with other deployment morphologies. The suburban and rural results are presented in Table~\ref{tab:flops_count}. In brief, the performance loss at median is at most $1.54\%$ for 32 and 64 APs, and at most $3.2\%$ for 128 APs. The general trend in all these results is that the GNN inference can be slightly degraded by bigger graphs and different deployment morphologies than the ones used for training. However, the performance remains close to the optimal for all practical purposes.

\begin{figure}
\centering
\includegraphics[width=0.99\linewidth]{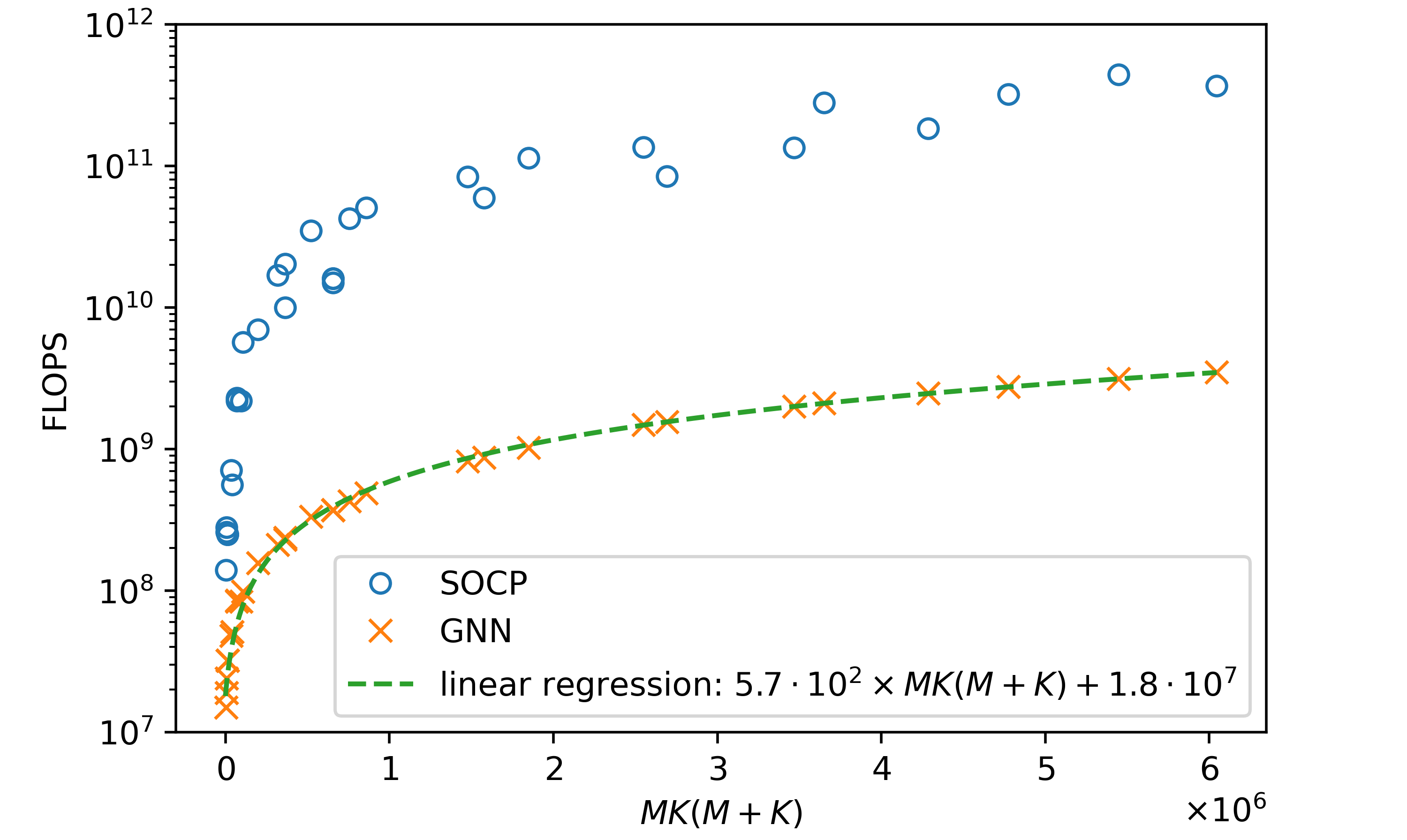}
\caption{FLOPs of GNN and SOCP as a function of $MK(M+K)$}\label{fig:flops_plot}
\vspace{-0.5cm}
\end{figure}

Let us discuss about the computational complexity of our solution. The graph structure defined in Section~\ref{sec:graph_repr} has $MK$ nodes, $MK(M-1)$ edges of type UE and $MK(K-1)$ edges of type AP. Since the GNN operations are computed on the graph's nodes and edges, one can show that its asymptotic computational complexity is $O\!\left(MK(M+K)\right)$.
To validate this by simulation, we run the algorithms on randomly generated large-scale fading realizations for different values of $M = 1,\ldots,256$ and $K=5,\ldots,72$, then we plot their FLOPs versus $MK(M+K)$ in Fig.~\ref{fig:flops_plot}. We see that the GNN's FLOPs can be well fitted by a linear function in $MK(M+K)$ which confirms the above result.

In Fig.~\ref{fig:flops_plot} and Table~\ref{tab:flops_count}, GNN has 10 times fewer FLOPs than SOCP on small scenarios with 32 APs. For 128 APs, it requires about 40 times fewer FLOPs than SOCP, and for 256 APs, it reduces the FLOPs by a factor 105. The same observation can be made on the run-times: on a CPU\textsuperscript{1}, the GNN computes the power control for 128 APs and 32 UEs in 1s, while it takes 45s for SOCP to complete. In real-world systems, the GNN would be implemented on a GPU: we observe under 50ms of run-times on a Nvidia TITAN RTX GPU.


{\renewcommand{\arraystretch}{1.2}
\begin{table}[t]
    \centering
    \vspace{0.3cm}
    \caption{Comparison of GNN and SOCP}\label{tab:flops_count}
    \resizebox{0.485\textwidth}{!}{
    \begin{tabular}{ |c|c|c|c|c| }
         \hline
         \multirow{2}{*}{Scenario} & \multicolumn{2}{c|}{FLOPs \vphantom{$\frac{1}{1}$}} 
         & GNN loss & GNN $95\%$
         \\ \cline{2-3}
         & GNN \vphantom{$\frac{1}{1}$} & SOCP & at median & likely loss\\
      \hline
      \cellcolor{gray!20} Urban: & \multicolumn{3}{c}{}\\
      \hline 
      24 APs, 5 UEs & \num{1.5e7} & \num{1.4e8} & $0.73\%$ & $1.42\%$\\
      \hline
      32 APs, 5 UEs & \num{1.9e7} & \num{2.8e8} & $0.48\%$ & $1.19\%$\\
      \hline
      32 APs, 6 UEs & \num{2.4e7} & \num{2.6e8} & $0.43\%$ & $0.51\%$\\
      \hline
      32 APs, 9 UEs & \num{3.2e7} & \num{2.5e8} & $0.47\%$ & $1.05\%$\\
      \hline
      48 APs, 12 UEs & \num{4.8e7} & \num{7.1e8} & $0.77\%$ & $1.51\%$\\
      \hline
      64 APs, 9 UEs & \num{5.1e7} & \num{5.6e8} & $0.19\%$ & $0.25\%$\\
      \hline
      64 APs, 18 UEs & \num{8.4e7} & \num{2.2e9} & $0.35\%$ & $0.57\%$\\
      \hline
      96 APs, 30 UEs & \num{2.3e8} & \num{1.0e10} & $1.56\%$ & $2.39\%$\\
      \hline
      128 APs, 32 UEs & \num{3.7e8} & \num{1.5e10} & $2.05\%$ & $2.94\%$\\
      \hline
      \cellcolor{gray!20} Suburban: & \multicolumn{3}{c}{}\\
      \hline
      32 APs, 9 UEs & \num{3.2e7} & \num{2.5e8} & $1.54\%$ & $5.27\%$\\
      \hline
      64 APs, 18 UEs & \num{8.5e7} & \num{2.2e9} & $1.20\%$ & $2.41\%$\\
      \hline
      128 APs, 32 UEs & \num{3.7e8} & \num{1.6e10} & $3.20\%$ & $4.28\%$\\
      \hline
      \cellcolor{gray!20} Rural: & \multicolumn{3}{c}{}\\
      \hline
      32 APs, 9 UEs & \num{3.2e7} & \num{2.5e8} & $1.35\%$ & $4.02\%$\\
      \hline
      64 APs, 18 UEs & \num{8.4e7} & \num{2.3e9} & $0.78\%$ & $1.88\%$\\
      \hline
      128 APs, 32 UEs & \num{3.7e8} & \num{1.6e10} & $2.97\%$ & $3.91\%$\\
      \hline
    \end{tabular}}
    \vspace{-0.3cm}
\end{table}
}

\section{Conclusion}\label{s:conc}
In this paper, we propose a GNN to solve the downlink max-min power control for a CFmMIMO system with MRT precoded beamforming. Our solution takes advantage of the problem's permutation equivariance property to greatly improve the learning efficiency and accuracy. Numerical results show that a single trained GNN achieves near-optimal performance for various systems sizes and deployment scenarios. Furthermore, its complexity remains low in all the simulated use-cases, therefore it is implementable in practice even in very large systems. The aforementioned two points demonstrate the superiority of our approach over the state of the art in terms of scalability and generalizability.

\bibliographystyle{IEEEtran}
\bibliography{IEEEabrv,reference}

\begin{thebibliography}{10}
\providecommand{\url}[1]{#1}
\csname url@samestyle\endcsname
\providecommand{\newblock}{\relax}
\providecommand{\bibinfo}[2]{#2}
\providecommand{\BIBentrySTDinterwordspacing}{\spaceskip=0pt\relax}
\providecommand{\BIBentryALTinterwordstretchfactor}{4}
\providecommand{\BIBentryALTinterwordspacing}{\spaceskip=\fontdimen2\font plus
\BIBentryALTinterwordstretchfactor\fontdimen3\font minus
  \fontdimen4\font\relax}
\providecommand{\BIBforeignlanguage}[2]{{%
\expandafter\ifx\csname l@#1\endcsname\relax
\typeout{** WARNING: IEEEtran.bst: No hyphenation pattern has been}%
\typeout{** loaded for the language `#1'. Using the pattern for}%
\typeout{** the default language instead.}%
\else
\language=\csname l@#1\endcsname
\fi
#2}}
\providecommand{\BIBdecl}{\relax}
\BIBdecl

\bibitem{marzetta2010}
T.~L. Marzetta, ``Noncooperative cellular wireless with unlimited numbers of
  base station antennas,'' \emph{{IEEE} Trans. Wireless Commun.}, vol.~9,
  no.~11, pp. 3590--3600, 2010.

\bibitem{ym2014}
H.~Yang and T.~L. Marzetta, ``A macro cellular wireless network with uniformly
  high user throughputs,'' in \emph{IEEE Veh. Technol. Conf.}, 2014.

\bibitem{ym2013}
------, ``Capacity performance of multicell large-scale antenna systems,'' in
  \emph{51st Annual Allerton Conference on Communication, Control, and
  Computing}, 2013.

\bibitem{ngo2017}
H.~Q. Ngo, A.~Ashikhmin, H.~Yang, E.~G. Larsson, and T.~L. Marzetta,
  ``Cell-free massive {MIMO} versus small cells,'' \emph{{IEEE} Trans. Wireless
  Commun.}, vol.~16, no.~3, pp. 1834--1850, 2017.

\bibitem{elina2017}
E.~Nayebi, A.~Ashikhmin, T.~L. Marzetta, H.~Yang, and B.~D. Rao, ``Precoding
  and power optimization in cell-free massive {MIMO} systems,'' \emph{{IEEE}
  Trans. Wireless Commun.}, vol.~16, no.~7, pp. 4445--4459, 2017.

\bibitem{zhang2020}
J.~Zhang, E.~Bj{\"o}rnson, M.~Matthaiou, D.~W.~K. Ng, H.~Yang, and D.~J. Love,
  ``Prospective multiple antenna technologies for beyond {5G},'' \emph{{IEEE}
  J. Sel. Areas Commun.}, vol.~38, no.~8, pp. 1637--1660, 2020.

\bibitem{carmen2019}
C.~D’Andrea, A.~Zappone, S.~Buzzi, and M.~Debbah, ``Uplink power control in
  cell-free massive {MIMO} via deep learning,'' in \emph{IEEE 8th International
  Workshop on Computational Advances in Multi-Sensor Adaptive Processing
  (CAMSAP)}, 2019.

\bibitem{zhang2021}
Y.~Zhang, J.~Zhang, Y.~Jin, S.~Buzzi, and B.~Ai, ``Deep learning-based power
  control for uplink cell-free massive {MIMO} systems,'' in \emph{IEEE
  Globecom}, 2021.

\bibitem{raja2021}
N.~Rajapaksha, K.~B.~S. Manosha, N.~Rajatheva, and M.~Latva-aho, ``Deep
  learning-based power control for cell-free massive {MIMO} networks,'' in
  \emph{IEEE International Conference on Communications}, 2021.

\bibitem{zhao2020}
Y.~Zhao, I.~G. Niemegeers, and S.~Heemstra~de Groot, ``Power allocation in
  cell-free massive {MIMO}: A deep learning method,'' \emph{{IEEE} Access},
  vol.~8, no.~5, pp. 87\,185--87\,200, 2020.

\bibitem{yan2020globecom}
H.~Yan, A.~Ashikhmin, and H.~Yang, ``Optimally supporting {IoT} with cell-free
  massive {MIMO},'' in \emph{IEEE Globecom}, 2020.

\bibitem{yan2020}
------, ``A scalable and energy efficient {IoT} system supported by cell-free
  massive {MIMO},'' \emph{{IEEE} Internet Things J.}, 2021.

\bibitem{salaun2021deep}
L.~Sala{\"u}n and H.~Yang, ``Deep learning based power control for cell-free
  massive {MIMO} with {MRT},'' in \emph{IEEE Globecom}, 2021.

\bibitem{luo2022}
L.~Luo, J.~Zhang, S.~Chen, X.~Zhang, B.~Ai, and D.~W.~K. Ng, ``Downlink power
  control for cell-free massive {MIMO} with deep reinforcement learning,''
  \emph{{IEEE} Trans. Veh. Technol.}, 2022, early Access.

\bibitem{scarselli2009}
F.~Scarselli, M.~Gori, A.~C. Tsoi, M.~Hagenbuchner, and G.~Monfardini, ``The
  graph neural network model,'' \emph{{IEEE} Trans. Neural Netw.}, vol.~20,
  no.~1, pp. 61--80, 2009.

\bibitem{keriven2019universal}
N.~Keriven and G.~Peyr{\'e}, ``Universal invariant and equivariant graph neural
  networks,'' \emph{Advances in Neural Information Processing Systems},
  vol.~32, 2019.

\bibitem{yang2018energy}
H.~Yang and T.~L. Marzetta, ``Energy efficiency of massive {MIMO}: cell-free
  vs. cellular,'' in \emph{IEEE 87th Veh. Technol. Conf.}, 2018.

\bibitem{ITU-R2009}
M.~Series, ``Guidelines for evaluation of radio interface technologies for
  {IMT-Advanced},'' \emph{Report ITU M.2135-1}, 2009.

\bibitem{ba2016layer}
J.~L. Ba, J.~R. Kiros, and G.~E. Hinton, ``Layer normalization,'' \emph{arXiv
  preprint arXiv:1607.06450}, 2016.

\bibitem{shi2020masked}
Y.~Shi, Z.~Huang, S.~Feng, H.~Zhong, W.~Wang, and Y.~Sun, ``Masked label
  prediction: Unified message passing model for semi-supervised
  classification,'' \emph{arXiv preprint arXiv:2009.03509}, 2020.

\bibitem{vaswani2017attention}
A.~Vaswani, N.~Shazeer, N.~Parmar, J.~Uszkoreit, L.~Jones, A.~N. Gomez,
  {\L}.~Kaiser, and I.~Polosukhin, ``Attention is all you need,''
  \emph{Advances in neural information processing systems}, vol.~30, 2017.

\bibitem{kingma2014adam}
D.~P. Kingma and J.~Ba, ``{Adam}: A method for stochastic optimization,''
  \emph{arXiv preprint arXiv:1412.6980}, 2014.

\bibitem{FeyLenssen2019}
M.~Fey and J.~E. Lenssen, ``Fast graph representation learning with {PyTorch
  Geometric},'' in \emph{ICLR Workshop on Representation Learning on Graphs and
  Manifolds}, 2019.

\bibitem{andersen2003implementing}
E.~D. Andersen, C.~Roos, and T.~Terlaky, ``On implementing a primal-dual
  interior-point method for conic quadratic optimization,'' \emph{Mathematical
  Programming}, vol.~95, no.~2, pp. 249--277, 2003.

\end{thebibliography}

\end{document}